\documentclass[english]{llncs}
\usepackage[latin1]{inputenc}
\usepackage[T1]{fontenc}
\usepackage{babel}
\usepackage{graphicx} 
\usepackage{amsmath}
\usepackage{amsfonts}
\usepackage{amssymb}
\usepackage{multirow}
\usepackage{multicol}
\usepackage{url}
\usepackage{stmaryrd}
\usepackage{lmodern}
\usepackage{longtable}
\usepackage{rotating}
\usepackage{array}
\usepackage{pdflscape}
\usepackage{soul} 
\usepackage{xspace}
\usepackage{enumitem}
\usepackage[normalem]{ulem}
\usepackage{proof}
\usepackage{hyperref}

\usepackage{tikz}
\usetikzlibrary{arrows, shapes, trees, positioning, decorations.markings, patterns}
\usetikzlibrary{chains,positioning}
\usepackage{bpmn-events}
\usepackage{bpmn-gateways}
\usepackage{bpmn-misc}

\usepackage{color}

\newcommand{\en}{\mathit{enacting}}

\newcommand{\seq}{\mathit{seq}}

\newcommand{\com}{\textit{completes}}
\newcommand{\beg}{\textit{begins}}
\newcommand{\tkn}{\textit{enables}}

\newcommand{\task}{\mathit{task}}

\newcommand{\er}{F}

\newcommand{\etr}{\longrightarrow}

\newcommand{\eq}{\textit{eq}}
\newcommand{\start}{\mathit{start}\xspace}
\newcommand{\eend}{\mathit{end}\xspace}

\newcommand{\paru}{\hspace{-.2mm}\raisebox{-2pt}{-}\hspace{-.2mm}}  
\newcommand{\excu}{\hspace{-.0mm}\raisebox{-2pt}{-}\hspace{-.2mm}}  

\newcommand{\pbr}{\textit{par\paru branch}} 
\newcommand{\pme}{\textit{par\paru merge}}  
\newcommand{\ebr}{\textit{exc\excu branch}} 
\newcommand{\eme}{\textit{exc\excu merge}}  
\newcommand{\upp}{\rule{0mm}{3.5mm}}

\newcommand{\verimap}{\textsc{VeriMAP}\xspace} 
\newcommand{\eldarica}{\textsc{Eldarica}\xspace} 
\title{Verification of Time-Aware Business Processes\\
using Constrained Horn Clauses}

\author{Emanuele De Angelis\inst{1} \and Fabio Fioravanti\inst{1} \and
Maria Chiara Meo\inst{1}\\ 
Alberto Pettorossi\inst{2} \and Maurizio Proietti\inst{3}}

\institute{DEC, University `G. D'Annunzio', Pescara, Italy\\
\texttt{\{emanuele.deangelis,fabio.fioravanti,cmeo\}@unich.it}\\
\and DICII, University of Rome Tor Vergata,
Rome, Italy\\
\texttt{pettorossi@disp.uniroma2.it}\\
\and IASI-CNR, Rome, Italy \ \
\texttt{maurizio.proietti@iasi.cnr.it}}

\begin{document}

\maketitle

\begin{abstract}
We present a method for verifying properties of 
time-aware business processes, that is, business process
where time constraints on the activities are explicitly
taken into account.
Business processes are specified using an
extension of the Business Process Modeling Notation (BPMN) and 
durations are defined by constraints over integer numbers.
The definition of the operational
semantics is given by a set {\it OpSem} of constrained 
Horn clauses (CHCs). 
Our verification method consists of two steps. (Step~1)~We  
specialize {\it OpSem} with respect to a given business process 
and a given temporal property to be verified, whereby getting
a set of CHCs whose satisfiability is equivalent to the 
validity of the given property. 
(Step~2)~We use state-of-the-art solvers for CHCs to check the satisfiability
of such sets of clauses.
We have implemented our verification method using the VeriMAP transformation
system, and the \eldarica and Z3 solvers for CHCs.
\end{abstract}

\section{Introduction}
\label{sec:intro}

A {\em business process}, or BP for short, consists of a set of activities,
performed in coordination within a single organization, 
which realize a business goal~\cite{Ho&10,Wes07}.
The Business Process Model and Notation, or BPMN for short,  
is one of the most popular graphical languages proposed for 
visualizing business processes~\cite{BPMN13}. 
The primary goal of BPMN is to provide a standard notation that can be understood 
by all business stakeholders, which  include the business analysts who 
define and modify the processes, the technical developers in charge of 
their implementation, 
and the business managers who monitor and manage them.

A BPMN model is a procedural, semi-formal description of the
order of execution of the activities of a given process 
and how these activities must coordinate,
abstracting away from many other aspects of the process itself, 
such as the manipulation of data
and the duration of the activities. However, for many analysis tasks 
these aspects are very significant in practice. In particular, the duration
of the activities is critical, when we want to reason about
time constraints (e.g., deadlines) that should be satisfied by process
executions.

Various approaches for BP modeling with duration and 
time constraints have been proposed in the literature
(see~\cite{Ch&15} for a recent survey).
Some of these approaches define the semantics of {\em time-aware} BPMN models
by means of formalisms such as {\em time Petri nets}~\cite{Ma&10},
{\em timed automata}~\cite{Wa&11}, and {\em process algebras}~\cite{WoG09}.
Properties of these models can then be verified by using very effective 
reasoning tools available for those formalisms~\cite{BeV06,FDR98,La&97}.
 
However, the above mentioned formalisms and tools may not be
adequate if we want to complement time-based reasoning
with general purpose logical reasoning,
which is often needed if we take into account more complex 
aspects of knowledge manipulation activities relative to business processes. For instance,
some verification approaches make use of ontology-based reasoning about 
the business domain where processes are executed~\cite{SmP13,We&10},
while others combine reasoning on the finite-state process behavior with reasoning on
the manipulation of data objects of an infinite type, such as databases or
integers~\cite{Da&12,Ba&13,PrS14}.

Thus, in view of an integration of various reasoning tasks needed
to analyze business processes from different perspectives,
we propose a logic-based approach to modeling and verifying 
time-aware business processes.

The main contributions of the paper are the following.
We present a logic-based language to specify time-aware BPMN models,
where time and duration of activities are explicitly represented.
Then we define an operational semantics of time-aware BPMN models
by means of deduction rules that allow us to infer the time intervals
when a particular activity is in execution or `enacting' (using the BPMN terminology).
Next, in order to prove properties of time-aware BPMN models, we follow a
transformational approach similar to the one proposed in~\cite{De&15b} 
for the verification
of imperative programs. First, we consider an encoding {\it OpSem} of 
the operational semantics into {\em Constrained Horn Clauses} (CHCs)~\cite{Bj&15} 
(or, equivalently, {\em Constraint Logic Programs}~\cite{JaM94}). 
Then, we specialize {\it OpSem} to the time-aware BPMN model under
consideration and temporal property of interest,
thereby deriving a new set of CHCs whose satisfiability is equivalent to 
(and thus implies) the 
validity of the property. 
Finally, we use state-of-the-art solvers for CHCs (in particular, 
\eldarica~\cite{Ho&12} and Z3~\cite{DeB08})
to check the satisfiability of such set of clauses.

Since the CHCs are generated in an automatic way by the
CHC specializer from the formal definition of the semantics of the BPMN models, 
and the CHC solvers are general purpose reasoning systems, our approach 
is, to a large extent, parametric with respect to other extensions of BP models
one may want to consider in the future. Moreover, recent advances in the field of
CHC solving can be exploited to get very effective reasoning tools
for verifying properties of business processes.

The paper is structured as follows.
In Section~\ref{sec:pre} we recall some basic notions about 
Constrained Horn Clauses over integer numbers and BPMN.
In Section~\ref{sec:sem} we present our logic-based language for specifying
time-aware BPMN models and the operational semantics of the language.
In Section~\ref{sec:specint} we present the CHC encoding
of the semantics and the transformation techniques for
specializing  {\it OpSem} with respect to a given time-aware BPMN model
and a given property.
In Section~\ref{sec:implementation} we report on the implementation
of the verification technique we have made using the VeriMAP transformation
and verification system~\cite{De&14b}, and the CHC solvers \eldarica and Z3.
Finally, in Section~\ref{sec:RW} we discuss related work in the
field of BP verification.

\section{Preliminaries}
\label{sec:pre}

In the next two subsections we recall some basic notions 
concerning constrained Horn clauses and the Business Process Model and Notation.

We consider discrete time and we model the time line as the set of integers.
However, our approach applies directly to dense or continuous time.

\subsection{Constrained Horn Clauses over Integers}
\label{sec:CLP}

First we need the following notions about constraints, 
constrained Horn clauses, and constraint logic programming. 
For related notions not familiar to the reader, 
we refer to~\cite{JaM94,Llo87}.

Constraints are defined as follows. 
Let $\textit{RelOp}$ be the set of predicate symbols $\{=,\neq,\leq,\geq,<,>\}$.
If $p_1$ and $p_2$ are linear polynomials with integer variables 
and coefficients, then $p_1 R\,p_2$, with $R\!\in\!\textit{RelOp}$, is an 
{\em atomic constraint}.
A {\em constraint} $c$ is a (possibly empty) conjunction of atomic constraints. 
An {\it atom} is a formula of the form $p(t_{1},\ldots,t_{m})$,
where $p$ is a predicate symbol not in $\textit{RelOp}$ and 
$\mathit{t_{1},\ldots,t_{m}}$ are terms constructed as usual from variables, constants, 
and function symbols. 
In particular, we assume that there are the two predicate symbols \textit{true} and \textit{false}
of arity 0, and a predicate symbol $\textit{eq}$ denoting identity.
A~{\it constrained Horn clause}  (or simply, a {\it clause}) is an implication of the form  
$A\leftarrow c, G$, where the conclusion (or {\it head\/}) $A$ is an atom, 
and the premise (or {\it body\/}) $c$ is a constraint, and~$G$ is a (possibly empty)
conjunction of atoms. The empty conjunction is identified with \textit{true}.
A~{\it constrained goal} (or simply, a {\it goal}\/) is a clause of the form  
$\textit{false} \leftarrow c, G$.

An {$\mathbb Z$}-{\em interpretation} of a set~$\mathcal P$ of CHCs is defined to be an 
interpretation~$I$ of $\mathcal P$ such that: 
(i) \textit{true} holds in $I$, (ii)~\textit{false} does not hold in $I$,
(iii) $I$ is the usual interpretation over the set of the {integer} numbers
{$\mathbb Z$} for the constraints,  and
(iv) $I$ is the Herbrand interpretation for predicate and function symbols not in 
$\textit{RelOp}\,\cup\{\textit{true},\textit{false},+,\times\}$
(in particular, $\textit{eq}(x,y)$ holds if and only if $x$ and $y$ are identical
terms in the Herbrand universe).
An {$\mathbb Z$}-{\em model} of $\mathcal P$ is an 
\mbox{{$\mathbb Z$}-{interpretation}}~$M$ such that 
every clause of $\mathcal P$ holds in $M$.
A set of CHCs is {\em satisfiable} if it has an \mbox{{$\mathbb Z$}-{\em model}}.
(Note that a set of CHCs may be unsatisfiable if it contains goals.)
Every satisfiable set~$\mathcal P$ of CHCs has a unique 
{\em least {$\mathbb Z$}-model},  denoted $M(\mathcal P)$~\cite{JaM94}.

\subsection{Business Processes Model and Notation}
\label{section:bpmn}

A BPMN model is defined through a diagram drawn
by using graphical constructs representing \textit{flow objects} and 
\textit{sequence flows} (sequence flows will also be called \textit{flows} for short).
That diagram
can be extended, if so desired, to include information about data flow, 
resource allocation (i.e., how the work to be done is 
assigned to the participants in the process), and exception handling 
(i.e., how erroneous behaviors should be handled). 

For reasons of simplicity, in this paper we will only consider a subset of 
the {flow objects} and {sequence flows} that can occur in a BPMN model, but 
our approach can easily be extended to full BPMN.
The flow objects we will consider 
are of three kinds only: either (i)~\textit{tasks}, denoted by rounded rectangles, 
or (ii)~\textit{events}, denoted by circles, or (iii)~\textit{gateways}, 
denoted by diamonds. 
Tasks represent atomic units of work performed within the process. 
Events denote something that `happens' during the enactment of a business process.
We will only consider \textit{start events} and \textit{end events}, which start 
and end the process enactment, respectively.
Gateways model the branching and merging of activities. 
There are several types of gateways in BPMN, each of which can be a \textit{branch} 
gateway if it has multiple outgoing flows and a single incoming flow, or 
a \textit{merge} gateway if it has multiple incoming flows and a single outgoing flow.  
We will consider the following gateways: 
(i) the \textit{parallel branch} gateway
that concurrently activates all the outgoing flows,
(ii) the \textit{parallel merge} gateway that activates the outgoing flow when all the
incoming flows have been activated (that is, the parallel merge
synchronizes the incoming flows)
(iii) the \textit{exclusive branch} gateway that (non-deterministically) activates 
exactly one out of many outgoing flows, and
(iv) the \textit{exclusive merge} gateway that activates the single outgoing flow 
upon activation of one of the incoming flows.
The diamonds representing parallel and exclusive gateways are labeled by
`$+$' and `$\times$', respectively. 
A sequence flow, denoted by an arrow,
links two flow objects and denotes a control flow relation, i.e., 
it states that the control flow can pass from the source to the target object.
If there is a sequence flow from $x$ to $y$, then $x$ is a {\em predecessor} of $y$
and $y$ is a {\em successor} of $x$.
A {\em path} in a BPMN model is a sequence of flow objects such that every pair of consecutive
objects is connected by a sequence flow.

We assume that BPMN models are \textit{well-formed},
that is, they satisfy the following properties:
(1)  every process contains a unique start event and a unique end event,
(2)  every flow object occurs on a path from the start event to the end event,
(3)  start events have exactly one successor and no predecessor, 
(4)  end events have exactly one predecessor and no successor,
(5)  branch gateways have exactly one predecessor and at least one successor, 
     while merge gateways have at least one predecessor and exactly one successor, 
(6)  tasks have exactly one predecessor and one successor, and
(7)  on every cyclic path there is at least one occurrence of a task 
     (i.e., no cycles through gateways only are allowed).

\begin{figure}[htbp]
\begin{center}
\resizebox{\textwidth}{!}{

\begin{tikzpicture}[
      line width = 1pt
      , auto
      , start chain = going right
      , node distance = 15mm
      , minimum size=8mm
      ]
        
\tikzstyle{arr}=[->,  thick]
        
  \node [StartEvent, radius=5mm, label={[yshift=2mm]below:{\it start}}] (start) {start};

  \node [ExclusiveGateway,draw,right of=start, label={[yshift=2mm]below:$g1$} ] (g1) {};
  \node [task, right of=g1, label={[yshift=2mm]below:$a$}] (a) {add\\item};
  \node [ExclusiveGateway,draw,right of=a,   label={[yshift=2mm]below:$g2$} ] (g2) {};
  \node [task, right of=g2, label={[yshift=2mm]below:$p$}] (p) {pay};
  \node [ParallelGateway,draw,right of=p,   label={[yshift=2mm]below:$g3$} ] (g3) {};
  \node [task, above right of=g3, label={[yshift=-2mm]above:$i$}] (i) {issue\\invoice};

  \node [task, below right of=g3, label={[yshift=2mm]below:$o$}] (o) {prepare\\order};
  \node [ExclusiveGateway,draw,right of=o,   label={[yshift=2mm]below:$g4$}] (g4) {};
  \node [task, above right of=g4, label={[yshift=2mm]below:$sd$}] (sd) {standard\\delivery};
  \node [task, below right of=g4, label={[yshift=2mm]below:$ed$}] (ed) {express\\delivery};
  \node [ExclusiveGateway,draw,below right of=sd,   label={[yshift=2mm]below:$g5$} ] (g5) {};
  \node [ParallelGateway,draw, above right of=g5,    label={[yshift=2mm]below:$g6$} ] (g6) {};

  \node [task, above left of=g6, label={[yshift=-2mm]above:$s$}] (s) {send\\invoice};

  \node [EndEvent,right of=g6, label={[yshift=2mm]below:{\it end}}] (end) {end};

 \draw [arr] (start) -- (g1);
 \draw [arr] (g1) -- (a);
 \draw [arr] (a) -- (g2);
 \draw [arr] (g2) -- ++(0,1.2) -| (g1);
 \draw [arr] (g2) -- (p);
 \draw [arr] (p) -- (g3);
 \draw [arr] (g3) -- (i) ;
 \draw [arr] (i) -- (s) ;
 \draw [arr] (s) -- (g6);
 \draw [arr] (g3) -- (o) ;
 \draw [arr] (o) -- (g4) ;
 \draw [arr] (g4) -- (sd) ;
 \draw [arr] (sd) -- (g5) ;
 \draw [arr] (g4) -- (ed) ;
 \draw [arr] (ed) -- (g5) ;
 \draw [arr] (g5) -- (g6);
 \draw [arr] (g6) -- (end);

%
%
%
\end{tikzpicture}
} 

\caption{The BPMN model of the purchase order process~$PO$.
The italicized labels are not part of the model and are only used to
denote the corresponding flow objects.}
\label{fig:ex}
\end{center}
\end{figure}
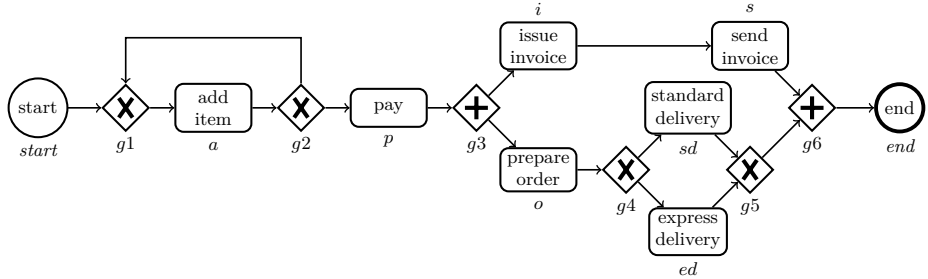

 In Figure~\ref{fig:ex} we show the BPMN model of 
 a purchase order process~$PO$,
 describing a typical interaction pattern between an e-commerce vendor and a customer.
 
At the beginning of the purchase order process 
 the customer adds one or more items to the shopping cart.
Subsequently, the customer pays for the items then
the vendor 
(i)~issues the invoice then sends it to the customer, and
(ii)~prepares the order then ships it using either a standard or an express delivery method. 
The process terminates when the invoice has been sent and the order has been delivered.

\section{Specification and Semantics of Business Processes}
\label{sec:sem}

In this section we introduce the notion of a {\em Business Process Specification} (BPS), 
which is a way of formally representing a business process by means of CHCs.
Then we define the operational semantics of a BPS.

\subsection{Specifying Business Processes through CHCs}
\label{subsec:BPS}

A BPS~$\mathcal B$ contains a set of ground {facts} of the 
form $p(c_1,\dots ,c_n)$, where $c_1,\dots ,c_n$ are constants denoting flow 
objects (that is, either tasks, or events, or gateways) and \textit{p} is a predicate
symbol. We will make use of the following predicates:

\noindent
- $\textit{flow\paru object}(x)$: $x$ is either a task, or an event, or a gateway;

\noindent
- $\task(x)$: $x$ is a task;

\noindent
- $\start(e)$ and $\eend(e)$: $e$ is a start event and an end event, respectively;

\noindent
- $\seq(x,y)$: there is a sequence flow from $x$ to $y$;

\noindent
- $\mathit{par\paru branch}(g)$ and $\mathit{par\paru merge}(g)$:
$g$ is a parallel branch and a parallel merge gateway, respectively;

\noindent
- $\mathit{exc\excu branch}(g)$ and $\mathit{exc\excu merge}(g)$: $g$ is an exclusive 
branch and exclusive merge gateway, respectively;

\noindent
- $\mathit{duration}(x,d)$: the enactment of the flow object $x$ takes $d$ units of time to be 
completed.

We assume that: (i) for every task~$x$ there exists in~$\mathcal B$ a single clause 
of the form $\textit{duration}(x,d)\!\leftarrow\!d_{min}\!\leq\!d\!\leq\!d_{max}$, where
$d_{min}$ and $d_{max}$ are positive integer constants representing the minimal 
and the maximal time duration of~$x$, respectively, and 
(ii) for every event and gateway $x$ there exists in $\mathcal B$ a single clause
of the form $\textit{duration}(x,d)\!\leftarrow\!d\!=\!0$ (that is,
events and gateways are instantaneous).

The CHC specification of the  BPMN process {\it PO} of Figure \ref{fig:ex}
is shown in Table~\ref{tab:enc}. In our {\it PO} example we will use
the standard Prolog syntax for clauses.

\begin{table}[htbp]
\vspace*{-5mm}
\begin{center}
{\tt \small

\begin{tabular}{l l l l l@{\hspace{22mm}} l@{\hspace{2mm}}}
{start(start).} &  {end(end).}  \\[2mm]
{exc\_merge(g1).} &  {exc\_branch(g2).}  &    {exc\_branch(g4).} &  {exc\_merge(g5).} & \\
{par\_branch(g3).}  &{par\_merge(g6).} \\[2mm]
\end{tabular}

\begin{tabular}{l @{\hspace{2mm}}l@{\hspace{2mm}} l@{\hspace{2mm}} l @{\hspace{2mm}}l @{\hspace{2mm}}l}
seq(start,g1). & seq(g2,g1). & seq(g3,i). & seq(g3,o). & seq(g4,sd). & seq(ed,g5). \\
seq(g1,a). & seq(g2,p). & seq(i,s). & seq(o,g4). & seq(g4,ed). & seq(g5,g6).\\
seq(a,g2). & seq(p,g3). & seq(s,g6). & 	seq(sd,g5). & seq(g6,end). \\[2mm]
\end{tabular}

\begin{tabular}{l@{\hspace{4mm}} l@{\hspace{4mm}} l @{\hspace{41mm}}l l l}
task(a). &  {duration(a, D):- D>=1, D=<6.} &    {\%\,add\,item}\\
task(p). &  {duration(p, D):- D>=1, D=<2.} &    {\%\,pay} \\
task(i). &  {duration(i, D):- D>=1, D=<2.} &    {\%\,issue\,invoice} \\
task(s). &  {duration(s, D):- D>=1, D=<3.} &    {\%\,send\,invoice} \\
task(o). &  {duration(o, D):- D>=3, D=<5.} &    {\%\,prepare\,order} \\
task(sd). &  {duration(sd,D):- D>=2, D=<4.}&    {\%\,deliver\,order\,(standard)} \\
task(ed). & {duration(ed,D):- D>=1, D=<3.}&     {\%\,deliver\,order\,(express)} \\
 & {duration(X, D):- not\_task(X), D=0.} &      {\%\,gateways\,and\,events}
\end{tabular}
}
\end{center}
\label{tab:enc}
\caption{The set of facts encoding the schema of the
purchase order process $PO$
of Figure~\ref{fig:ex}. }
\vspace*{-5mm}
\end{table}


Note that a BPS is always satisfiable (note that, in particular, it contains no goals),
and hence it has a least {$\mathbb Z$}-model.

Our formalization also includes a set $\mathcal W$ of clauses that represent
the {\em meta-model} of the BPS, defining: (i) disjointness relationships among 
sets of elements, 
for instance, $\textit{false} \leftarrow \task(X), \pbr(X)$,
(ii) properties of the BPS corresponding to Conditions 1--7
of Section~\ref{section:bpmn}, which define 
the \textit{well-formedness} of a~BPMN model.
These properties are expressed as CHCs as follows:


\begin{enumerate}
\item[(1)] \makebox[52mm][l]{$\eq(X,Y)\!\leftarrow\!\start(X), \start(Y)$}
and ~ $\eq(X,Y)\!\leftarrow\!\eend(X), \eend(Y)$;

\item[(2)] \makebox[56mm][l]{$\textit{seq}^{*}(S,X)\!\leftarrow\!\start(S),$ 
$\textit{flow\paru object}(X)$} and\, $\textit{seq}^{*}(X,E)\!\leftarrow\!\textit{flow\paru object}(X), 
\eend(E)$

\hspace{10mm}where ${\mathit{seq}}^{*}$ is the reflexive, transitive closure of $\seq$;

\item[(3)] \makebox[62mm][l]{$\eq(Y,Z)\!\leftarrow\!\start(S),\seq(S,Y),\seq(S,Z)$}
 and ~ $\textit{false}\!\leftarrow\!\start(S),$ $\seq(Y,S)$;

\item[(4)] \makebox[62mm][l]{$\eq(Y,Z)\!\leftarrow\!\eend(E),\seq(Y,E),\seq(Z,E)$}
 and ~ $\textit{false}\!\leftarrow\!\eend(E),$ $\seq(E,Y)$;

\item[(5)] $\eq(Y,Z)\!\leftarrow\!\pbr(X), \seq(Y,X), \seq(Z,X)$ \ and \ 

$\eq(Y,Z)\!\leftarrow\!\pme(X),$ $\seq(X,Y),$ $\seq(X,Z)$ 

\hspace{10mm} and, similarly, for the \ebr  ~and \eme ~gateways;

\item[(6)] $\eq(Y,Z)\!\leftarrow\!\task(X), \seq(X,Y), \seq(X,Z)$ \  and

$\eq(Y,Z)\!\leftarrow\!\task(X), \seq(Y,X), \seq(Z,X)$;

\item[(7)] $\textit{false}\!\leftarrow\!\textit{gateway} \excu \textit{path}(X,X)$

\hspace{10mm} 
\hangindent=10mm where $\textit{gateway} \excu \textit{path}(X,Y)$ is a predicate 
that holds iff there is a path\\[-4mm]

\hspace{10mm} from $X$ to $Y$ with gateways only.

\end{enumerate}\vspace{-2mm}

Note that the existence
of at least one predecessor and at least one successor 
for any task or gateway (required by Conditions 5 and 6 of
Section~\ref{section:bpmn}) is enforced by the clauses at Point 2.

A BPS $\mathcal B$ is well-formed if all clauses in $\mathcal W$ hold in the
least $\mathbb Z$-model of $\mathcal B$.

\subsection{Operational Semantics}
\label{subsec:opsem}

We start off by introducing the notion of a {\it state}, which is represented
by a set of properties, called {\em fluents}, that hold at a given time point.
A state $s\in\mathit{States}$ is a pair $\langle F, t \rangle$, where 
$F$ is a set of {fluents} and $t$ is a time point in {$\mathbb Z$}. 

A fluent is a term of one of the following forms:
(i)~$\beg(x)$, which represents the beginning of the enactment (or execution) of the
flow object~$x$,
(ii)~$\com(x)$, which represents that~$x$ has completed its execution, and
(iii)~$\tkn(x,y)$, which represents that the flow object~$x$ has completed
its execution and it enables the execution of its successor~$y$, and
(iv)~$\en(x,r)$, which represents that the enactment of~$x$ 
requires~$r$ units of time to completion (for this reason 
$r$ is also called the \textit{residual time} of $x$).
Thus, $\beg(x)$ is equivalent to $\en(x,r)$, where $r$ is the duration of~$x$,
and $\com(e)$ is equivalent to $\en(x,0)$. (This redundancy of representation
allows us to write simpler rules for the operational semantics below.)

The operational semantics is defined by a binary 
transition relation $\etr$ which is a subset of $\mathit{States}\times\mathit{States}$. 
The {\em initial state}, denoted {\it init}, is the pair 
$\langle\{\beg(start)\},0\rangle$.
In the rules below, which define $\etr$, 
we also use the following predicates, besides the ones
introduced in Section \ref{subsec:BPS}:
(i) $\mathit{not}\paru\pbr(x)$, which holds if $x$ is not a parallel branch, and
(ii) $\mathit{not}\paru\pme(x)$, which holds if $x$ is not a parallel merge.

\vspace{2mm}
\noindent
\raisebox{3.5mm}{$(S_{1})$}~~~\infer[]{\upp \langle\er,t\rangle \etr 
\langle ( \er  \setminus \{
  \beg(x)  
\} )
\cup \{
  \en(x,d)
\},\ t \rangle
}{
     \beg(x) \!\in\!\er    \hspace{12mm}     \mathit{duration}(x,d)
} 

\vspace{2mm}


\noindent

\vspace{2mm}
\noindent
\raisebox{3.5mm}{$(S_{2})$}~~~\infer[]{\upp \langle\er,t\rangle \etr 
\langle (\er \setminus \{
  \com(x)  
\} )
\cup \{
  \tkn(x,s)\ |\  \seq(x,s)
\},\ t \rangle
}{
    \com(x) \!\in\!\er 		 \hspace{14mm}    \mathit{par\paru branch}(x)
}

\vspace{2mm}
\noindent

\vspace{2mm}
\noindent
\raisebox{3.5mm}{$(S_{3})$}~~~\infer[]{\upp \langle\er,t\rangle \etr 
\langle ( \er  \setminus \{
  \com(x)  
\} )
\cup \{
  \tkn(x,s)
\},\ t \rangle
}{
    ~~\com(x)  \!\in\!\er  \hspace{10mm} 		\textit{not\excu par\paru branch}(x) \hspace{10mm} \seq(x,s)~~
}

\vspace{2mm}

\noindent

\vspace{2mm}
\noindent
\raisebox{3.5mm}{$(S_{4})$}~~~\infer[]{\upp \langle\er,t\rangle \etr 
\langle (\er  \setminus   \{\tkn(p,x)~|~\tkn(p,x)\in \er \}   )
\cup \{
  \beg(x) 
\},\ t\rangle
}{
  \forall p ~ \mathit{seq}(p,x) \rightarrow \tkn(p,x)\in\er
  \hspace{12mm}   \mathit{par\paru merge}(x)
}


\vspace{2mm}
\noindent	

\vspace{2mm}
\noindent
\raisebox{3.5mm}{$(S_{5})$}~~~\infer[]{\upp \langle\er,t\rangle \etr 
\langle ( \er  \setminus \{
  \tkn(p,x)  
\} )
\cup \{
  \beg(x)
\},\ t \rangle
}{
    \tkn(p,x) \!\in\!\er \hspace{8mm}   \mathit{not\paru par\paru merge}(x)
}


\vspace{2mm}
\noindent	

\vspace{2mm}
\noindent\raisebox{3.5mm}{$(S_{6})$}~~~\infer[]{\upp \langle\er,t\rangle \etr 
\langle ( \er  \setminus \{
   \en(x,0) 
\} )
\cup \{
  \com(x)
\},\ t \rangle
}{
     \en(x,0) \!\in\!\er }


\vspace{2mm}
\noindent


\vspace{2mm}
\noindent
\raisebox{3.5mm}{$(S_{7})$}~~~\infer[]{\upp \makebox[110mm]{$\langle\er,t\rangle \etr	
\langle(\er \setminus \{\en(x,r)~|~\en(x,r) \in\er \})$} }{\hspace{-7mm}
{\mathit{no\paru other\paru premises}}(F) \hspace{12mm} \exists x\, \exists r\, \en(x,r) 
\!\in\!\er \hspace{12mm} m\!>\!0 
}

\hspace*{31.8mm}$\cup~ \{\en(x,r\!-\!m)~|~\en(x,r)\in\er\},\ t\!+\!\textit{m}\rangle
$

\vspace{2mm}
\hangindent=6mm
where: (i)~${\mathit{no\excu other\paru premises}}(F)$ holds iff none of the 
rules $S_{1}$--$S_{6}$ has its premise true, 
and 
(ii)~$\textit{m} = min\{r~|~\en(x,r)\!\in\!\er\}$.

\medskip
Let us first observe that $S_7$ is the only rule that formalizes
the flow of time, as it infers transitions of the form
$\langle F, t \rangle \longrightarrow \langle F', t\!+\!m \rangle$, with $m\!>\!0$.
In contrast, rules $S_1$--$S_6$ infer 
instantaneous state transitions, that is, transitions of the form
$\langle F, t \rangle \longrightarrow \langle F', t \rangle$.

Rules $S_1$--$S_7$ have the following meaning.\vspace*{-2mm}

\begin{itemize}
\item[($S_1$)] If the execution of a flow element $x$ begins at time~$t$, then, 
at the same time~$t$,
$x$ is enacting and its residual time is the duration $d$ of $x$;

\item[($S_2$)] If the execution of the parallel branch $x$ completes at 
time $t$, then $x$ enables {\em all its successors} at time $t$;

\item[($S_3$)] If the execution of $x$ completes at time~$t$
and~$x$ is not a parallel branch, then~$x$ enables {\em precisely one 
of its successors} at time $t$ {(in particular, this case occurs when 
 $x$ is a task);}

\item[($S_4$)]  If {\em all} the predecessors of $x$ have enabled the parallel merge $x$ at time $t$,
then the execution of $x$ begins at time $t$;

\item[($S_5$)] If {\em at least one} predecessor $p$ of $x$ enables $x$ at time $t$ and
$x$ is not a parallel merge, then the execution of $x$ begins at time $t$
{(in particular, this case occurs when $x$ is a task);}

\item[($S_6$)] If a flow object $x$ is enacting at time $t$ with residual time 0, 
then the execution of $x$ completes at time $t$;

\item[($S_7$)] Suppose that: (i)~none of rules $S_1$--$S_6$ is applicable to infer the
successor state of $\langle F,t\rangle$, (ii)~at time $t$ 
at least one task is enacting with {\em positive} residual time (note that
flow objects different from tasks cannot have positive residual time), and (iii)~$m$ 
is the least among the residual times of all the tasks enacting at time $t$.
Then every task $x$ that is enacting at time $t$ with residual time $r$,
is enacting at time $t+m$ with residual time $r-m$.
\end{itemize}

\noindent
We say that state  $\langle F',t'\rangle$ is {\em reachable}
from state $\langle F,t\rangle$, if
$\langle F,t\rangle \longrightarrow^* \langle F',t'\rangle$, where
$\longrightarrow^*$ denotes the reflexive, transitive closure of the
transition relation~$\etr$.

\section{Encoding Time-Dependent Properties of Business Processes into CHCs}
\label{sec:specint}

In this section we show the CHC \textit{interpreter} that encodes 
the operational semantics and the property to be verified.
We also present two transformation techniques:
(1) a technique for
\textit{removing the interpreter} and
deriving a set of clauses 
that is amenable to automatic satisfiability checking,
and 
(2) a technique for reducing the size of sets of CHC clauses 
by using a suitable notion of predicate equivalence.

\subsection{Encoding the Operational Semantics in CHCs}
\label{subsec:CHCsemantics-1}

A state $\langle\er,t\rangle$ 
of the operational semantics 
is encoded by a term of the form \texttt{s(F,T)},
where \texttt{F} is a list encoding the set $\er$ of fluents and \texttt{T} encodes the time point $t$
at which the fluents in the set $F$ hold.
The transition relation $\longrightarrow$ between states
and its reflexive, transitive closure $\longrightarrow^{*}$
are encoded by the binary
predicates $\texttt{tr}$ and  $\texttt{reach}$, respectively,
whose defining clauses are shown in Table~\ref{fig:interpreter}.
In the body of the clauses, the atoms that encode 
the premises of the rules of the operational semantics 
have been underlined.

The predicate \texttt{member(X,L)} selects an element \texttt{X} from the list \texttt{L}.
The predicate
\texttt{update(F,R,A,FU)} holds iff \texttt{FU} is the list obtained from the list \texttt{F}
by removing 
all the elements of~\texttt{R}  
and adding 
all the elements of~\texttt{A}. 
The predicate \texttt{no\_other\_premises(F)} holds iff the premise of every rule in 
$\{{\mathtt{S1}},\ldots, {\mathtt{S6}}\}$ is false.
The predicate \texttt{mintime(Enacts,M)} holds iff \texttt{Enacts} is a list of terms of the form \texttt{enacting(X,R)}
and \texttt{M} is the minimum value of \texttt{R} for the elements of \texttt{Enacts}.
The predicate \texttt{decrease\_residual\_times(Enacts,M,EnactsU)} holds iff \texttt{EnactsU} is the list of terms 
obtained by replacing every element of \texttt{Enacts}, of the form \texttt{enacting(X,R)}, 
with the term \texttt{enacting(X,RU)} where \texttt{RU\,=\,R-M}. 
The predicates \texttt{sublist(S,L)} and \texttt{findall(X,G,L)} have the usual meaning.

\vspace{-3mm}
\begin{table}[ht]
{\small


\noindent    %
\hspace{0cm}\makebox[6mm][l]{\tt S1.}\texttt{tr(s(F,T), s(FU,T))} {\tt{:- }} \underline{\texttt{member(begins(X),F)}}\!{\tt,} \underline{\texttt{duration(X,D)}}\!{\tt,}

\vspace{0mm}\hspace{40mm}
\texttt{update(F,[begins(X)],[enacting(X,D)],FU). }

\noindent    %
\hspace{0cm}\makebox[6mm][l]{\tt S2.}\texttt{tr(s(F,T), s(FU,T))} {\tt{:- }} \underline{\texttt{member(completes(X),F)}}\!{\tt,} \underline{\texttt{par\_branch(X)}}\!{\tt,}

\vspace{0mm}\hspace{40mm}
\texttt{findall(enables(X,S),(seq(X,S)),Enbls),}

\vspace{0mm}\hspace{40mm}
\texttt{update(F,[completes(X)],Enbls,FU). }

\noindent    %
\hspace{0cm}\makebox[6mm][l]{\tt S3.}\texttt{tr(s(F,T), s(FU,T))} {\tt{:-}} \underline{\texttt{member(completes(X),F)}}\!{\tt,}\,\underline{\texttt{not\_par\_branch(X),seq(X,S)}}\!{\tt,}

\vspace{0mm}\hspace{40mm}
\texttt{update(F,[completes(X)],[enables(X,S)],FU).}

\noindent    %
\hspace{0cm}\makebox[6mm][l]{\tt S4.}\texttt{tr(s(F,T), s(FU,T))} {\tt{:- }} \underline{\texttt{member(enables(\_,\_),F)}}\!{\tt,} \underline{\texttt{par\_merge(X)}}\!{\tt,}

\vspace{0mm}\hspace{40mm}
\underline{\texttt{findall(enables(P,X),(seq(P,X)),Enbls)}}\!{\tt,}

\vspace{0mm}\hspace{40mm}
\texttt{\underline{sublist(Enbls,F),} update(F,Enbls,[begins(X)],FU).}

\noindent    %
\hspace{0cm}\makebox[6mm][l]{\tt S5.}\texttt{tr(s(F,T), s(FU,T))} {\tt{:- }} \underline{\texttt{member(enables(P,X),F)}}\!{\tt,} \underline{\texttt{not\_par\_merge(X)}}\!{\tt,}

\vspace{0mm}\hspace{40mm}
\texttt{update(F,[enables(P,X)],[begins(X)],FU). }

\noindent    %
\hspace{0cm}\makebox[6mm][l]{\tt S6.}\texttt{tr(s(F,T), s(FU,T))} {\tt{:- }} \underline{\texttt{member(enacting(X,R),F)}}\!{\tt,} \underline{\texttt{R=0}}\!{\tt,}

\vspace{0mm}\hspace{40mm}
\texttt{update(F,[enacting(X,R)],[completes(X)],FU). }

\noindent    %
\hspace{0cm}\makebox[6mm][l]{\tt S7.}\texttt{tr(s(F,T), s(FU,TU))} {\tt{:- }} \underline{\texttt{no\_other\_premises(F)}}\!{\tt,}\ \underline{\texttt{member(enacting(\_,\_),F)}}\!{\tt,}

\vspace{0mm}\hspace{40mm}
\underline{\texttt{findall(Y,(Y=enacting(X,R),member(Y,F)),Enacts)}}\!{\tt,}

\vspace{0mm}\hspace{40mm}
\underline{\texttt{mintime(Enacts,M)}}\!{\tt,}\ \underline{\texttt{M>0}}\!{\tt,}

\vspace{0mm}\hspace{40mm}
\texttt{decrease\_residual\_times(Enacts,M,EnactsU),}

\vspace{0mm}\hspace{40mm}
\texttt{update(F,Enacts,EnactsU,FU), TU=T+M. }

\vspace{2mm}\noindent       %
\hspace{0mm}\makebox[6mm][l]{\tt R1.}$\texttt{reach(S,S).}$ \hspace{8mm} {\tt R2.} $\texttt{reach(S,S2)}$ {\tt{:-}} $\texttt{tr(S,S1), reach(S1,S2).}$


} 

\vspace{2mm}
\caption{\label{table:ms1}The CHC interpreter for the 
operational semantics of time-aware business processes.}
\label{fig:interpreter}
\vspace{-10mm}
\end{table}

\subsection{Encoding Time-Dependent Properties}
\label{subsec:Removal}

By using the {\tt reach} predicate and integer constraints, we can
specify many interesting time-dependent properties.
In particular, we can specify safety properties 
(stating that `no unsafe state can be reached'),
schedulability properties (stating that a process will be completed
within a given deadline), response properties (stating that, whenever
a task is executed, another task will be executed within a given time),
and many other quantitative temporal properties.

In order to see how we encode time-dependent properties of business processes,
we consider a property of the $PO$ process 
stating that, whenever the customer pays and the process completes, then completion
occurs within~9 time units from payment.
By using the reachability relation, this property can be written as follows: 

\smallskip

\noindent
\ \ \textit{prop}:
if $\textit{init} \rightarrow^{*} \langle \{\textit{completes}(p)\},t_p \rangle  
\rightarrow^{*} \langle \{\textit{completes}(e)\},t_e \rangle $,  
then $t_e\leq t_p+9$.

\smallskip

\noindent
The reader can check that \textit{prop} holds for the $PO$ process 
because, in the worst case, 
the time needed for preparing and delivering the order is actually 9 time units
and this time is greater than the time needed for issuing and sending the invoice, 
which is 5 time units.
The property {\it prop} is encoded by the following goal:

\smallskip
\noindent    %
{\small
\hspace{0cm}\makebox[6mm][l]{\tt NP.}\texttt{false} {\tt{:- }} \texttt{Ts=0, Tp>Ts, Te>Tp+9,}

\vspace{0mm}\hspace{15.1mm}
\texttt{reach(s([begins(start)],Ts), s([completes(p)],Tp)),}

\vspace{0mm}\hspace{15.1mm}
\texttt{reach(s([completes(p)],Tp),  s([completes(e)],Te)).}
}

\smallskip
\noindent
The clauses \texttt{S1-S7,R1,R2,NP},
together with the clauses encoding the $PO$ process,
will be collectively referred to as the \textit{interpreter} $I$.
We have that the property \textit{prop} is valid for the {\it PO} process
iff the set  $I$ of CHCs is satisfiable.

Despite several tools have been developed for checking the satisfiability of constrained
Horn clauses, 
none of them can effectively be leveraged in our example.
Constraint logic programming systems~\cite{JaM94} are focused on
proving the unsatisfiability of sets of clauses, rather then their satisfiability,
and they fail to terminate
for the given set $I$ because of recursive {\tt reach} clause (note, in particular
that the {\tt add\_item} task can be executed an unbounded number of times).
State-of-the-art CHC solvers~\cite{DeB08,Ho&12} also fail
because the predicates in $I$ are defined over lists and structured terms (not just integers) 
and they depend on the \texttt{findall} predicate, 
which is not available in those solvers.

In order to be able to effectively use off-the-shelf CHC solvers 
for checking the validity of time-dependent properties,
we apply the so-called \textit{removal of the interpreter}
transformation~\cite{De&15b,Pe&98},
a program specialization strategy based on unfold/fold transformation rules, 
which takes the program $I$ 
as input
and produces as output a program $I_\textit{sp}$
that is equivalent to $I$ with respect to satisfiability.
Indeed, by the correctness of the unfold/fold transformation rules~\cite{EtG96}, we have that
$I$ is satisfiable
iff
$I_\textit{sp}$ is satisfiable.

A notable effect of applying the {removal of the interpreter} 
is that the program $I_\textit{sp}$
contains no occurrences of the predicates and terms
used for encoding the operational semantics and the $PO$ process.
Indeed, the clauses of $I_\textit{sp}$ will be of the form
$A \leftarrow c, B$, where the arguments of the atoms are variables
and $c$ is a constraint. For instance, in the {\it PO}
example, the goal expressing the property {\it prop}
is transformed into the goal:
\smallskip

{\small
\texttt{false} {\tt{:- }} \texttt{A=0, B=<2, C=<6, D=<5, E>0, F-E>9, B>=1, C>=1, D>=3,}

\vspace{0mm}\hspace{12.4mm}
\texttt{ new1(C,A,E), new2(B,D,E,F).}
}

\smallskip
\noindent
The new predicates {\tt new1} and {\tt new2} have been 
introduced by the {\em definition rule}, and the extra constraints have been derived
by the {\em unfolding rule}. 
We refer to~\cite{De&15b} for the details of the transformation.
The whole set of clauses derived by the removal of the interpreter
is listed in Appendix A.1. The satisfiability of this set of clauses 
can be proved in a fully automatic way by using either the \textsc{Eldarica}\xspace or
the Z3 solver, 
as it will be demonstrated in Section~\ref{sec:implementation}.

\subsection{Predicate Equivalence}
\label{subsec:Post}

Now we introduce a transformation that allows us to
reduce the size of a set of CHC clauses when suitable
equivalences between predicates hold.
Since  predicate equivalence is undecidable in general,
we consider a decidable notion of predicate equivalence based on 
predicate names and constraint equivalence.

We assume, without loss of generality,
that all clauses are in {\em pure} form, that is,
of the form $p(X) \leftarrow B$, where $X$ is a tuple of distinct variables.
Let $P$ be a set of CHCs. By $\mathit{Pred}(P)$ we denote the set of predicate 
symbols occurring in $P$. A {\em predicate renaming} for $P$ is a,
{possibly not injective,} mapping 
$\pi\!: \mathit{Pred}(P) \rightarrow Q$, where $Q$ is a set of predicate
symbols. Given a set $S$ of formulas with
predicates in $\mathit{Pred}(P)$,
$\pi(S)$~is a new set of formulas obtained by replacing,
for all predicates $p\!\in \!\mathit{Pred}(P)$,  every occurrence
of $p$ in $S$ by $\pi(p)$.

For every $k$-ary predicate $p\! \in\! \mathit{Pred}(P)$ we assume that
all clauses for $p$ have head $p(X)$, where $X$ is a $k$-tuple of
distinct variables.
We define
$\mathit{Bodies}(p(X),P)$ to be the set $ \{B ~|~ p(X) \leftarrow B 
\mbox{ is a clause in } P\}$.
We write 
$\mathit{Bodies}(p(X),P) \equiv \mathit{Bodies}(q(X),P)$
if there exists a bijection 
\mbox{$\eta\! : \mathit{Bodies}(p(X),P) \rightarrow \mathit{Bodies}(q(X),P)$}
such that, for every $B \!\in\! \mathit{Bodies}(p(X),P)$, 
$\exists Y\, B$ and $\exists Z\, \eta(B)$ are equivalent modulo constraints,
where $Y$ is the tuple of variables occurring in $B$ and not in $X$,
and  $Z$ is the tuple of variables occurring in $\eta(B)$ and not in $X$.

\begin{definition}[Predicate Equivalence]\label{def:equiv}
Let $P$ be a set of clauses in pure form, and $E = \{P_1,\ldots,P_n\}$ 
be a partition of $\mathit{Pred}(P)$.
For $i=1,\ldots,n,$ let $e_i$ be a predicate symbol in $P_i$,
and $\pi\! : \mathit{Pred}(P) \rightarrow \{e_1,\ldots,e_n\}$ be a predicate
renaming for $P$ such that, for $i=1,\ldots,n,$  $\pi(p)\!=\!e_i$ iff $p\! \in\! P_i$.

The partition~$E$ is a {\em cp-equivalence} on $P$ if, for $i=1,...,n,$ 
given any two predicates  $p,q$ in~$P_i$, 
$p$ and $q$ have the same arity $k$ and, for any $k$-tuple $X$ of
new, distinct variables, $\pi(\mathit{Bodies}(p(X),P)) \equiv \pi(\mathit{Bodies}(q(X),P))$.
\end{definition}

Note that one can compute the coarsest 
\mbox{cp-equi}\-val\-ence on $P$ by a greatest fixpoint construction
starting from the partition where all predicate symbols belong to the same 
equivalence class.

Given a cp-equivalence $E$ on $P$ together with
the predicate renaming $\pi$ considered in Definition~\ref{def:equiv}, 
we can transform $P$ into a
set $\widetilde \pi(P,E)$ of clauses in two steps:
(i)~we remove from $P$ all clauses whose head predicate does not appear in the
range of $\pi$, and
(ii)~we apply $\pi$ to the remaining clauses.

\begin{theorem}
For any cp-equivalence $E$ on a set  $P$ of clauses,
$P$ is satisfiable iff $\widetilde \pi(P,E)$ is satisfiable.
\end{theorem}

To see an example of cp-equivalence, let us consider the following subset
of the clauses derived by the removal of the interpreter
in the {\it PO} example:

\begin{verbatim}
new5(A,B,C,D) :- A=0, new21(B,C,D).
new5(A,B,C,D) :- A=0, B=0, E=<3, E>=1, new10(E,C,D).
new5(A,B,C,D) :- B=0, E=<3, E>=1, new7(A,E,C,D).
new5(A,B,C,D) :- E=0, F=-A+B, G=A+C, A-B=<0, A>0, new5(E,F,G,D).
new5(A,B,C,D) :- E=0, F=A-B, G=B+C, B>0, A-B>=0, new5(F,E,G,D).
new4(A,B,C,D) :- A=0, new21(B,C,D).
new4(A,B,C,D) :- A=0, B=0, E=<3, E>=1, new10(E,C,D).
new4(A,B,C,D) :- B=0, E=<3, E>=1, new6(A,E,C,D).
new4(A,B,C,D) :- E=0, F=-A+B, G=A+C, A-B=<0, A>0, new4(E,F,G,D).
new4(A,B,C,D) :- E=0, F=A-B, G=B+C, B>0, A-B>=0, new4(F,E,G,D).
\end{verbatim}

The following partition of the set of predicates occurring in the above
clauses is a cp-equivalence:

\smallskip
\noindent
$E = \mathtt {\{\{new5,new4\},\{new7,new6\},\{new21\},\{new10\} \}}$

\smallskip

\noindent
associated with the following predicate renaming: 

\smallskip

$\mathtt{\pi(new5)=\pi(new4)=new4}$

$\mathtt{\pi(new7)=\pi(new6)=new6}$

$\mathtt{\pi(new21)=new21}$ 

$\mathtt{\pi(new10)=new10}$ 

\smallskip

\noindent
By the transformation $\widetilde \pi$, the clauses for $\mathtt{new5}$ 
are removed and all occurrences of $\mathtt{new5}$
are replaced by $\mathtt{new4}$.
In Appendix A.2, we show the effect of this transformation
on the whole set of clauses derived by the removal of the interpreter.

\section{Automated Verification}
\label{sec:implementation}

We have implemented the transformation strategies presented in Section~\ref{subsec:Removal}
(\textit{Removal of the Interpreter}) and Section~\ref{subsec:Post} 
(\textit{Predicate Equivalence}) by using the \verimap transformation 
and verification system~\cite{De&14b}. 
Then, we have used the SMT solvers \eldarica\footnote{v1.2-rc in client-server mode 
with options {\tt-horn} {\tt-hsmt} {\tt-princess}} and Z3\footnote{v4.4.2, master 
branch as of 2016-02-18, with the Duality fixed-point engine. See:\\
\url{http://research.microsoft.com/en-us/projects/duality/default.aspx}} for checking 
the satisfiability of the CHCs generated by \verimap.
The satisfiability check requires the following two steps: 
(i)~a preliminary \textit{Translate} step, 
in which \verimap
translates the CHCs into the SMT-LIB language, and (ii)~the \textit{Verify} step, 
in which an SMT solver is invoked for checking satisfiability.

Now we report on the results
obtained by using our prototype implementation on the Purchase Order business 
process shown in Figure~\ref{fig:ex}. The experiments have been performed on an 
Intel Core i5-2467M 1.60GHz processor with 4GB of memory under GNU/Linux OS.
The removal of the interpreter (see Table~\ref{fig:interpreter}), that is, its
specialization with respect
to the facts encoding the business process (see Table~\ref{tab:enc}) and the temporal 
property (see clause \texttt{NP}) requires 0.42 seconds and generates a set 
${\mathit{RI}}$ 
of 51 clauses.
The transformation of the clauses~${\mathit{RI}}$ based on predicate equivalence 
requires 0.02 seconds and generates a set~${\mathit{PE}}$ of 33 clauses. 
Running the SMT solvers on the clauses~${\mathit{RI}}$ requires: 
(i)~1.28 seconds using \eldarica (0.11 seconds for \textit{Translate} and 1.17 seconds for \textit{Verify}) and 
(ii)~1.09 seconds using Z3 (0.12 seconds for \textit{Translate} and 0.97 seconds for \textit{Verify}).
Running the SMT solvers on ${\mathit{PE}}$  
requires: 
(i)~0.81 seconds using \eldarica (0.11 seconds for \textit{Translate} and 0.70 seconds for \textit{Verify}) and 
(ii)~0.68 seconds using Z3 (0.11 seconds for \textit{Translate} and 0.57 seconds for \textit{Verify}).
We have that both SMT solvers \eldarica and Z3 are able to prove the satisfiability of 
${\mathit{RI}}$ and~${\mathit{PE}}$. 
We may observe that the transformation times are negligible and, in particular, 
that the transformation based on predicate equivalence, by reducing the sizes of the sets of 
CHCs clauses, allows solvers to improve their performance.
Indeed, for both solvers, the difference between the 
\textit{Verify} time taken on 
${\mathit{RI}}$ and the one taken on ${\mathit{PE}}$
(that is, before and after the application of the transformation) is 
much higher than the time taken for applying the transformation itself.

\section{Related Work}
\label{sec:RW}

Several papers have proposed approaches to model business processes
with time constraints and, in particular, duration
\cite{Ar&08,Co&09,Fo&08,Ga&09,WoG09}  (see also~\cite{Ch&15} for a recent survey).

The approach of Arbab et al.~\cite{Ar&08} provides a translation of
BPMN into the coordination language REO. 
Due to REO's
Constraint Automata semantics, in principle this translation permits formal reasoning 
about BPMN processes depending on time and resources.
However, the paper does not provide any formalized verification technique.

The workflow conceptual
model proposed in~\cite{Co&09} enables 
the specification and analysis of time constraints in business processes.
The paper proposes temporal constructs to express duration,
delays, relative, absolute, and periodic constraints. 
They also introduce the concept of controllability for workflow
schemata and its evaluation at process design time. Controllability refers to the
capability of executing a workflow for any possible duration of tasks,
where the minimum and the maximum durations for each task are known.
Their algorithms for testing controllability enumerate the possible
choices, and therefore suffers from memory growth.

Gonzalez del Foyo and Silva
consider in~\cite{Fo&08} workflow diagrams extended with task durations and the latest
execution deadline of each task. They provide a translation into Time Petri Nets~\cite{Be&91}, 
where clocks are associated with each transition in the net, and use the
tool TINA~\cite{BeV06} to answer schedulability questions.

The approach described in \cite{Ga&09} enables the specification of 
temporal constraints (such as `As Soon as Possible') and temporal dependencies.
However, unlike the approach presented here, no automated verification 
mechanism of time-dependent properties is provided

The approach presented in~\cite{WoG09} uses a timed semantic function which 
takes a diagram describing a collaboration, 
and returns a CSP process~\cite{Hoa78}  that models the
timed behavior of that diagram, by using the notion of a relative time in the form 
of delays chosen non-deterministically within given intervals.
Properties are then verified by using the FDR system~\cite{FDR98}.
Due to some intricacy of CSP,
some behavioral properties of business processes,
may not be easy to express for BP developers. 

Some proposals, 
such as~\cite{Wa&11} and others surveyed in~\cite{Ch&15},
make use of timed automata to model business processes with time constraints,
and use the UPPAAL tool~\cite{La&97} for the automated verification of some of
their properties.
As already mentioned, these proposals, as well as the ones cited above,
may not be adequate when taking into consideration properties
of business processes that require more advanced logical reasoning. 

Finally, we would like to mention work on modeling and analyzing business processes
with explicit time representation
based on the {\em Event Calculus}~\cite{KoS86}  (see, for instance,~\cite{Mo&14}).
However, the Event Calculus lacks a simple translation into constrained Horn clauses (in particular, it makes use of negation), which has been proposed in this paper as a means to 
enable the use of very effective automated verification systems.

\section{Conclusions}
\label{sec:Conclusions}

We have presented a logic-based language to specify BPMN models where time and 
duration of activities are explicitly represented.
The language enables the specification of timing constraints, given in
the form of lower and upper bounds associated with the duration of tasks.
These are useful features with an intuitive meaning that enable the specifier to
annotate activities with timing restrictions. 
The language supports the specification of a wide range of time-dependent properties,
such as the schedulability and response time.

The main advantage of our approach is that it allows us to automatically
generate constrained Horn clauses from 
the formal definition of the semantics of the BPMN models and the 
time-dependent properties of interest.
Then, by exploiting recent advances in the field of
CHC solving, we get very effective reasoning tools
for verifying properties of business processes.
Finally, the fact that our approach is parametric
with respect to the semantics of the process modeling languages we consider,
allows us to take into account future extensions of those
languages with very little effort.

\appendix

\newpage

\section{Output of transformations}

\subsection{Removal of the interpreter (Section \ref{subsec:Removal})}

{\footnotesize 
\begin{verbatim}
new44(A,B,C) :- A=0, B=C.
new44(A,B,C) :- D=0, E=A+B, A>0, new44(D,E,C).
new37(A,B,C) :- A=0, D=<3, D>=1, new17(D,B,C).
new37(A,B,C) :- A=0, D=<4, D>=2, new11(D,B,C).
new37(A,B,C) :- D=0, E=A+B, A>0, new37(D,E,C).
new21(A,B,C) :- A=0, D=<3, D>=1, new10(D,B,C).
new21(A,B,C) :- D=0, E=A+B, A>0, new21(D,E,C).
new17(A,B,C) :- A=0, B=C.
new17(A,B,C) :- D=0, E=A+B, A>0, new17(D,E,C).
new11(A,B,C) :- A=0, B=C.
new11(A,B,C) :- D=0, E=A+B, A>0, new11(D,E,C).
new10(A,B,C) :- A=0, B=C.
new10(A,B,C) :- D=0, E=A+B, A>0, new10(D,E,C).
new7(A,B,C,D) :- B=0, A=0, C=D.
new7(A,B,C,D) :- A=0, new10(B,C,D).
new7(A,B,C,D) :- B=0, new11(A,C,D).
new7(A,B,C,D) :- E=0, F=-A+B, G=A+C, A-B=<0, A>0, new7(E,F,G,D).
new7(A,B,C,D) :- E=0, F=A-B, G=B+C, B>0, A-B>=0, new7(F,E,G,D).
new6(A,B,C,D) :- B=0, A=0, D=C.
new6(A,B,C,D) :- A=0, new10(B,C,D).
new6(A,B,C,D) :- B=0, new17(A,C,D).
new6(A,B,C,D) :- E=0, F=-A+B, G=A+C, A-B=<0, A>0, new6(E,F,G,D).
new6(A,B,C,D) :- E=0, F=A-B, G=B+C, B>0, A-B>=0, new6(F,E,G,D).
new5(A,B,C,D) :- A=0, new21(B,C,D).
new5(A,B,C,D) :- A=0, B=0, E=<3, E>=1, new10(E,C,D).
new5(A,B,C,D) :- B=0, E=<3, E>=1, new7(A,E,C,D).
new5(A,B,C,D) :- E=0, F=-A+B, G=A+C, A-B=<0, A>0, new5(E,F,G,D).
new5(A,B,C,D) :- E=0, F=A-B, G=B+C, B>0, A-B>=0, new5(F,E,G,D).
new4(A,B,C,D) :- A=0, new21(B,C,D).
new4(A,B,C,D) :- A=0, B=0, E=<3, E>=1, new10(E,C,D).
new4(A,B,C,D) :- B=0, E=<3, E>=1, new6(A,E,C,D).
new4(A,B,C,D) :- E=0, F=-A+B, G=A+C, A-B=<0, A>0, new4(E,F,G,D).
new4(A,B,C,D) :- E=0, F=A-B, G=B+C, B>0, A-B>=0, new4(F,E,G,D).
new3(A,B,C,D) :- A=0, E=<3, E>=1, new6(E,B,C,D).
new3(A,B,C,D) :- A=0, E=<4, E>=2, new7(E,B,C,D).
new3(A,B,C,D) :- A=0, B=0, E=<3, E>=1, new17(E,C,D).
new3(A,B,C,D) :- A=0, B=0, E=<4, E>=2, new11(E,C,D).
new3(A,B,C,D) :- B=0, new37(A,C,D).
new3(A,B,C,D) :- E=0, F=-A+B, G=A+C, A-B=<0, A>0, new3(E,F,G,D).
new3(A,B,C,D) :- E=0, F=A-B, G=B+C, B>0, A-B>=0, new3(F,E,G,D).
new2(A,B,C,D) :- A=0, E=<3, E>=1, new3(B,E,C,D).
new2(A,B,C,D) :- B=0, E=<3, E>=1, new4(E,A,C,D).
new2(A,B,C,D) :- B=0, E=<4, E>=2, new5(E,A,C,D).
new2(A,B,C,D) :- A=0, B=0, E=<3, F=<3, E>=1, F>=1, new6(F,E,C,D).
new2(A,B,C,D) :- A=0, B=0, E=<3, F=<4, E>=1, F>=2, new7(F,E,C,D).
new2(A,B,C,D) :- E=0, F=-A+B, G=A+C, A-B=<0, A>0, new2(E,F,G,D).
new2(A,B,C,D) :- E=0, F=A-B, G=B+C, B>0, A-B>=0, new2(F,E,G,D).
new1(A,B,C) :- A=0, D=< 6, D>=1, new1(D,B,C).
new1(A,B,C) :- A=0, D=<2, D>= 1, new44(D,B,C).
new1(A,B,C) :- D=0, E=A+B, A>0, new1(D,E,C).
false :- A=0, B=<2, C=<6, D=<5, E>0, F-E>9, B>=1, C>=1, D>=3, new1(C,A,E), 
      new2(B,D,E,F).
\end{verbatim}
}

\subsection{Transformation Based on Predicate Equivalence (Section \ref{subsec:Post})}

The following partition of the set of predicates occurring in the 
clauses shown in A.1 is a cp-equivalence:

\smallskip
\noindent
$E = \mathtt {\{\{new44,new17,new11,new10\},\{new7,new6\},\{new5,new4\},\{new37\},\{new21\},\{new3\},}$

\hspace{4mm}$\mathtt {\{new2\},\{new1\}\}}$

\smallskip

\noindent
associated with the following predicate renaming: 

\smallskip

$\mathtt{\pi(new44)=\pi(new17)=\pi(new11)=\pi(new10)=new10}$

$\mathtt{\pi(new7)=\pi(new6)=new6}$

$\mathtt{\pi(new5)=\pi(new4)=new4}$

\smallskip

\noindent
and $\mathtt{\pi(p)=p}$ for all other predicate symbols.

\noindent
By applying $\widetilde \pi$, the clauses in A.1  are transformed into the following set:

{\footnotesize 
	\begin{verbatim}
	new37(A,B,C) :- A=0, D=<3, D>=1, new10(D,B,C).
	new37(A,B,C) :- A=0, D=<4, D>=2, new10(D,B,C).
	new37(A,B,C) :- D=0, E=A+B, A>0, new37(D,E,C).
	new21(A,B,C) :- A=0, D=<3, D>=1, new10(D,B,C).
	new21(A,B,C) :- D=0, E=A+B, A>0, new21(D,E,C).
	new10(A,B,C) :- A=0, B=C.
	new10(A,B,C) :- D=0, E=A+B, A>0, new10(D,E,C).
	new6(A,B,C,D) :- B=0, A=0, D=C.
	new6(A,B,C,D) :- A=0, new10(B,C,D).
	new6(A,B,C,D) :- B=0, new10(A,C,D).
	new6(A,B,C,D) :- E=0, F=-A+B, G=A+C, A-B=<0, A>0, new6(E,F,G,D).
	new6(A,B,C,D) :- E=0, F=A-B, G=B+C, B>0, A-B>=0, new6(F,E,G,D).
	new4(A,B,C,D) :- A=0, new21(B,C,D).
	new4(A,B,C,D) :- A=0, B=0, E=<3, E>=1, new10(E,C,D).
	new4(A,B,C,D) :- B=0, E=<3, E>=1, new6(A,E,C,D).
	new4(A,B,C,D) :- E=0, F=-A+B, G=A+C, A-B=<0, A>0, new4(E,F,G,D).
	new4(A,B,C,D) :- E=0, F=A-B, G=B+C, B>0, A-B>=0, new4(F,E,G,D).
	new3(A,B,C,D) :- A=0, E=<3, E>=1, new6(E,B,C,D).
	new3(A,B,C,D) :- A=0, E=<4, E>=2, new6(E,B,C,D).
	new3(A,B,C,D) :- A=0, B=0, E=<3, E>=1, new10(E,C,D).
	new3(A,B,C,D) :- A=0, B=0, E=<4, E>=2, new10(E,C,D).
	new3(A,B,C,D) :- B=0, new37(A,C,D).
	new3(A,B,C,D) :- E=0, F=-A+B, G=A+C, A-B=<0, A>0, new3(E,F,G,D).
	new3(A,B,C,D) :- E=0, F=A-B, G=B+C, B>0, A-B>=0, new3(F,E,G,D).
	new2(A,B,C,D) :- A=0, E=<3, E>=1, new3(B,E,C,D).
	new2(A,B,C,D) :- B=0, E=<3, E>=1, new4(E,A,C,D).
	new2(A,B,C,D) :- B=0, E=<4, E>=2, new4(E,A,C,D).
	new2(A,B,C,D) :- A=0, B=0, E=<3, F=<3, E>=1, F>=1, new6(F,E,C,D).
	new2(A,B,C,D) :- A=0, B=0, E=<3, F=<4, E>=1, F>=2, new6(F,E,C,D).
	new2(A,B,C,D) :- E=0, F=-A+B, G=A+C, A-B=<0, A>0, new2(E,F,G,D).
	new2(A,B,C,D) :- E=0, F=A-B, G=B+C, B>0, A-B>=0, new2(F,E,G,D).
	new1(A,B,C) :- A=0, D=<6, D>=1, new1(D,B,C).
	new1(A,B,C) :- A=0, D=<2, D>=1, new10(D,B,C).			
	new1(A,B,C) :- D=0, E=A+B, A>0, new1(D,E,C).
	false :- A=0, B=<2, D=<5, E>0, F-E>9, B>=1, C>=1, D>=3, C=<6, G=E, new1(C,A,G), 
	      new2(B,D,E,F).
	\end{verbatim}
}

\end{document}